\documentclass[journal]{IEEEtran}
\ifCLASSINFOpdf
\else
\fi

\usepackage{dirtytalk}
\usepackage{footnote}
\usepackage{times}
\usepackage{arydshln}
\usepackage{epsfig}
\usepackage{graphicx}
\usepackage[dvipsnames]{xcolor}
\usepackage{amsmath}
\usepackage{amssymb}
\usepackage{soul}
\usepackage{algorithm}
\usepackage{algorithmic}
\usepackage{footnote}
\usepackage{scrextend}
\usepackage{dirtytalk}
\usepackage{graphicx}
\usepackage{amsmath}
\usepackage{xspace}
\usepackage{adjustbox}
\usepackage{subcaption}
\usepackage{tablefootnote}
\usepackage{bm}
\usepackage{pifont}

\newcommand{\mypara}[1]{\vspace{2pt}\noindent{\bf{#1}}}

\newcommand{\Audiocaps}{\textsc{AudioCaps}\xspace}
\newcommand{\Clotho}{\textsc{Clotho}\xspace}
\newcommand{\QuerYD}{\textsc{QuerYD}\xspace}

\newcommand{\activityNetLong}{\textsc{ActivityNet-Captions}\xspace}
\newcommand{\activityNetShort}{\textsc{ActNetCaps}\xspace}
\newcommand{\datasetName}{\textsc{SoundDescs}\xspace}
\newcommand{\task}{text-audio retrieval\xspace}

\newcommand{\rev}[1]{\textcolor{black}{#1}}
\newcommand{\revv}[1]{\textcolor{black}{#1}}

\usepackage[breaklinks=true,colorlinks,bookmarks=false]{hyperref}
\usepackage{booktabs}

\begin{document}
\title{Audio Retrieval with Natural Language Queries: A Benchmark Study}

\author{
 A. Sophia Koepke$^{*}$\thanks{* Equal contribution.},
 Andreea-Maria~Oncescu$^{*}$, Jo\~{a}o~F.~Henriques, 
 Zeynep~Akata, and~Samuel~Albanie
 \thanks{A. S. Koepke and Z. Akata are with the Explainable Machine Learning group at the University of T{\"u}bingen. Z. Akata is also affiliated with the Max Planck Institute for Intelligent Systems, T{\"u}bingen and the Max Planck Institute for Informatics, Saarbr{\"u}cken.}
 \thanks{A.-M. Oncescu and J. Henriques are with the Visual Geometry Group at
 the University of Oxford.}
 \thanks{
 S. Albanie is with the Department of Engineering at the University of Cambridge.}

 }

\markboth{Journal of \LaTeX\ Class Files,~Vol.~14, No.~8, August~2015}%
{Shell \MakeLowercase{\textit{et al.}}: Bare Demo of IEEEtran.cls for IEEE Journals}

\maketitle

\begin{abstract}
The objectives of this work are \textit{cross-modal text-audio and audio-text retrieval},
in which the goal is to retrieve the audio content from a
pool of candidates that best matches a given written description and vice versa.
Text-audio retrieval enables users to search large databases
through an intuitive interface: they simply issue free-form
natural language descriptions of the sound they would like to hear.
To study the tasks of text-audio and audio-text retrieval,
which have received limited attention in the existing literature,
we introduce three challenging new benchmarks.
We first construct text-audio and audio-text retrieval benchmarks from
the \Audiocaps and \Clotho audio captioning datasets.
Additionally, we introduce the \datasetName benchmark,
which consists of paired audio and natural language
descriptions for a diverse collection of sounds that are
complementary to those found in \Audiocaps and \Clotho.
We employ these three benchmarks to establish baselines
for cross-modal text-audio and audio-text retrieval,
where we demonstrate the benefits
of pre-training on diverse audio tasks.
We hope that our benchmarks will inspire further research
into audio retrieval with free-form text queries. Code, audio features for all datasets used, and the \datasetName dataset are publicly available at \url{https://github.com/akoepke/audio-retrieval-benchmark}.
\end{abstract}


\IEEEpeerreviewmaketitle

\section{Introduction}
\label{sec:intro}

\IEEEPARstart{T}{he} vast and unabated growth of user-generated
content in recent years has introduced a pressing need to search
ever-growing databases of multimedia.
Free-form natural language sentences
(i.e. sequences of text that are written as they would be spoken)
form an intuitive and powerful interface for composing search queries
for these databases since they allow for expressing virtually any concept.
Spanning multiple modalities, different retrieval strategies
were developed for content as diverse as text (including web pages
and books), images~\cite{dong2016word2visualvec},
and videos~\cite{miech2018learning,mithun2018learning}.
Surprisingly, while search engines currently exist for these modalities
(e.g. Google, Flickr and YouTube, respectively),
unstructured audio is not accessible in the same way.
The aim of this paper is to address this gap by curating the
\datasetName dataset which contains paired sound and
natural language and by introducing benchmarks for
text-audio retrieval.

It is important to distinguish content-based retrieval
from that based on metadata,
such as the title of a video or song or an audio tag.
Metadata retrieval is feasible for manually-curated databases
such as song or movie catalogues.
However, content-based retrieval is more important in
user-generated data, which often has little structure or metadata.
There are methods to search for audio which matches
an audio query \cite{Manocha18,lallemand2012content},
but satisfying the requirement to input an example audio
query can be difficult for a human
(e.g. making convincing frog sounds is difficult).
We, on the other hand, propose a framework which enables
the searching of a sound database using detailed free-form
natural language \revv{queries} of the desired sound
\revv{(e.g. ``A man talking as music is playing followed by a frog croaking.'')}.
This enables the retrieval of audio data which will
ideally match the temporal sequence of events in the
query instead of just a single class tag.
Furthermore, natural language queries are a familiar
user interface widely used in current search engines.
Therefore, our proposed audio retrieval with free-form
text queries could be a first step towards more natural
and flexible audio-only search.

Text-based audio retrieval could also be beneficial
for video retrieval.
The majority of recent works that address the text-based
video retrieval task focus heavily on the visual and text 
domains~\cite{dong2016word2visualvec,miech2018learning,Liu19a,gabeur2020multi}.
Since audio and visual information inherently have
natural semantic alignment for a significant portion of
video data, text-based audio retrieval could also be used
for querying video databases by only considering the audio
stream of the video data.
This would allow for video retrieval in the audio domain
at reduced computational cost for cases in which audio
and visual information correspond, with applications
for low-power IoT devices, such as microphones in
natural habitats, of particular interest for conservation and biology.
Historical archives with extensive sound collections,
such as the British Library Sounds\footnote{\url{https://sounds.bl.uk}},
would be easier to search, facilitating historical research and public access.
Furthermore, text-based retrieval could enable appealing
creative applications, such as automatically finding (non-music) background sounds
which correspond to input text.
This could be especially useful given the growing popularity
of audio podcasts and audiobooks which are often supplemented
with (background) sounds that match their content.

Learning to retrieve audio, given natural language queries,
requires data with paired text and sound.
Audio captioning datasets naturally lend themselves to this task,
since they contain audio and a matching text description for the sound. 
However, existing captioning datasets are limited in size
and in the diversity of their audio content.
Hence, we curated the novel \datasetName dataset,
which was sourced from the BBC Sound Effects 
database\footnote{\url{https://sound-effects.bbcrewind.co.uk/}\label{bbc-footnote}}.
\datasetName contains text describing the sounds with significant variation with respect to the audio content and \rev2{with a relatively large} vocabulary used in the descriptions.

We introduce three new benchmarks for text-based audio retrieval,
based on our proposed \datasetName dataset,
and the \Audiocaps~\cite{kim2019audiocaps}
and \Clotho~\cite{drossos2020clotho} audio captioning datasets.
\Audiocaps consists of a subset of 10-second audio clips from the
AudioSet dataset~\cite{gemmeke2017audio}
with additional human-written audio captions,
while \Clotho contains sounds
sourced from the Freesound platform~\cite{font2013freesound},
varying between 15 and 30 seconds in duration, and accompanied by crowd-sourced text captions.
\datasetName is considerably more varied in duration and audio
content than the other two benchmarks.
However, the text descriptions are of mixed quality, since they are obtained
automatically from descriptions provided with the data.

In contrast to sound event class labels,
audio captions contain detailed information about the sounds.
A user searching for a particular sound would usually describe
the sound using text similar to an audio caption.
\Audiocaps~\cite{kim2019audiocaps}, \Clotho~\cite{drossos2020clotho},
and \datasetName allow to leverage the matching text-audio pairs
to train text-based audio retrieval frameworks.
To establish baselines for this task,
we adapt existing video retrieval frameworks for audio retrieval.
We employ multiple pre-trained audio expert networks and show that
using an ensemble of audio experts improves audio retrieval.

In summary, we make three contributions:
(1) we introduce the \datasetName dataset for text-based audio retrieval;
(2) we introduce three new benchmarks for audio retrieval
with natural language queries---to the best of our knowledge,
these represent the first public benchmarks for this task;
(3) we provide baseline performances with existing multi-modal
video retrieval models that we adapt to text-based audio retrieval
and show the benefits of using multiple datasets for pre-training.

This paper extends an initial Interspeech 2021 conference version
of our work~\cite{Oncescu21a} in two ways:
(i) we introduce the new \datasetName dataset for text-audio retrieval
and provide an analysis of its characteristics (in Sec.~\ref{sec:dataset}), 
(ii) we provide more extensive baselines for the audio retrieval task
with more detailed ablations across datasets and an additional
retrieval architecture (Sec.~\ref{sec:experiments}).
In particular, we explore the use of the Multi-modal Transformer
architecture~\cite{gabeur2020multi}.

\section{Related Work} \label{sec:related}
Our work relates to several themes in the literature: 
\textit{sound event recognition}, \textit{audio captioning}, \textit{audio-based retrieval}, \textit{text-based video retrieval} and \textit{text-domain audio retrieval}.
We discuss each of these next.

\mypara{Sound event recognition.}
There is a rich literature addressing the task of sound event recognition, 
which seeks to assign a given segment of audio with a corresponding semantic label.
Examples include detecting audio events associated with sports~\cite{xiong2003audio},
urban sounds \cite{aucouturier2007bag},
and distinguishing vocal and nonvocal events \cite{atrey2006audio}.
Research in this area has been driven by challenges,
such as DCASE~\cite{stowell2015detection,mesaros2017dcase},
and by the collection of sound event datasets.
These include TUT Acoustic scenes \cite{mesaros_annamaria_2017_400515},
CHIME-Home~\cite{foster2015chime},
ESC-50~\cite{piczak2015esc},
\mbox{FSD}Kaggle~\cite{fonseca2019audio},
and AudioSet~\cite{gemmeke2017audio}.
Of relevance to our approach, a number of prior works have employed deep learning for audio comprehension~\cite{Kong18,yu2018multi,kong2019weakly,ford2019deep,kong2020panns}.
Our work differs from theirs in that we focus instead on the task of
retrieval with natural language queries, rather than audio recognition.

\mypara{Audio captioning.}
Audio captioning consists of generating a natural language description for a sound~\cite{drossos2017automated}.
This requires a more detailed understanding of the sound than simply
mapping the sound to a set of labels (sound event recognition).
Recently, several audio captioning datasets have been introduced,
such as \Clotho~\cite{drossos2020clotho} which was used in the DCASE
automated audio captioning challenge 2020~\cite{Dcase20}, Audio Caption \cite{wu2019audio},
and \Audiocaps~\cite{kim2019audiocaps}.
Drawing inspiration from work on video captioning~\cite{venugopalan2015sequence,gao2017video}, 
multiple works have addressed automatic audio captioning on the \Audiocaps and \Clotho datasets \cite{koizumi2020audio,xu2021investigating,eren2020audio,mei2021audio,liu2021cl4ac}.
In this work, we use the \Audiocaps and \Clotho datasets for cross-modal retrieval.

\mypara{Audio-based retrieval.}
Multiple content-based audio retrieval frameworks, in particular query by example methods, leverage the similarity of sound features that represent different aspects of sounds (e.g. pitch, or loudness) \cite{foote1997content, Wold96, helen2007query, lallemand2012content}.
More recently, \cite{Manocha18} use a \revv{twin} neural network framework to
learn to encode semantically similar audio close together in the embedding space.
\cite{jin2012event} address multimedia event detection using only audio data,
while \cite{avgoustinakis2020audio} tackle near-duplicate video retrieval by audio retrieval.
These are purely audio-based methods that are applied to video datasets, but without using visual information. 
\cite{hou2013audio} propose a two-step approach for video retrieval which uses audio (coarse) and visual (fine) information together.

\mypara{Text-based video retrieval.}
More closely related to our work,
a number of methods showed that embedding video and text jointly into a shared space
(such that their similarity can be computed efficiently) is an effective approach~\cite{dong2016word2visualvec,mithun2018learning,miech2018learning,Liu19a,wray2019fine,gabeur2020multi,croitoru2021teachtext}
(though other formulations,
such as computing similarities directly in visual space
have also been explored~\cite{dong2018predicting}).
One particular trend has been to combine cues from
several ``experts''---pre-trained models that specialise
in different tasks (such as object recognition, action classification etc.)
to inform the joint embedding.
Recently, transformer-based architectures have demonstrated impressive results for text-based video retrieval~\cite{gabeur2020multi,bain2021frozen,luo2021clip4clip}. 
In this work, we propose to adapt three expert-based methods:
the Mixture of Embedded Experts method of~\cite{miech2018learning},
the Collaborative Experts model of~\cite{Liu19a},
and the Multi-Modal Transformer~\cite{gabeur2020multi}
by re-purposing them for the task of audio retrieval
(described in more detail in Sec.~\ref{sec:method}).

\mypara{Cross-domain audio retrieval.}
Methods that retrieve audio by matching associated text, such as metadata or sound event labels, have the implicit assumption that the text is relevant \cite{elizalde2018nels}.
\rev2{In contrast,} \cite{slaney2002semantic} is an early work that proposes to
link audio and text representations in hierarchical semantic and acoustic spaces. \cite{slaney2002mixtures} builds on this using mixture-of-probability-expert models for each of the modalities.
Chechik et al.~\cite{chechik2008large} propose a text-based
sound retrieval framework which uses single-word audio tags
as queries rather than caption-like natural language. Similarly, \cite{ikawa2018acoustic, ikawa2018generating} learn shared latent spaces between onomatopoeias (words that mimic non-speech sounds) and sound for searching audio using onomatopoeia queries and for generating sound words from audio.
The creative approach of \cite{aytar2017see} learns to align visual, audio, and text representations to enable cross-modal retrieval.
Their framework is trained with captioned images and paired image-sound data (sourced from videos) and evaluated using the soundtrack of captioned videos.
Other works have explored using images \cite{harwath2018jointly} or video data \cite{yasuda2020crossmodal,suris2018cross,zeng2020deep,jin2021cross,Nagrani18c} as queries for retrieving audio.
More recently,
\cite{elizalde2019cross} use a \revv{twin} network to learn a shared latent
text and sound space for cross-modal retrieval.
While they use class labels as text labels,
we study unconstrained text descriptions as queries.
Another highly related, concurrent line of works has explored
the task of grounding sounds given a text description.
\cite{xu2021text,tang2021query} presented results for the grounding task on
the AudioGrounding dataset, proposed by \cite{xu2021text},
which augments a subset of the \Audiocaps dataset
(approximately 10\% of the full \Audiocaps dataset)
by adding fine-grained temporal grounding labels.
In contrast to the audio grounding task,
text-based audio retrieval does not require expensive temporal annotations
and we can therefore leverage large databases which contain immensely varied content.

\section{\datasetName dataset} \label{sec:dataset}
\begin{figure}
    \centering
    \includegraphics[width=0.43\textwidth]{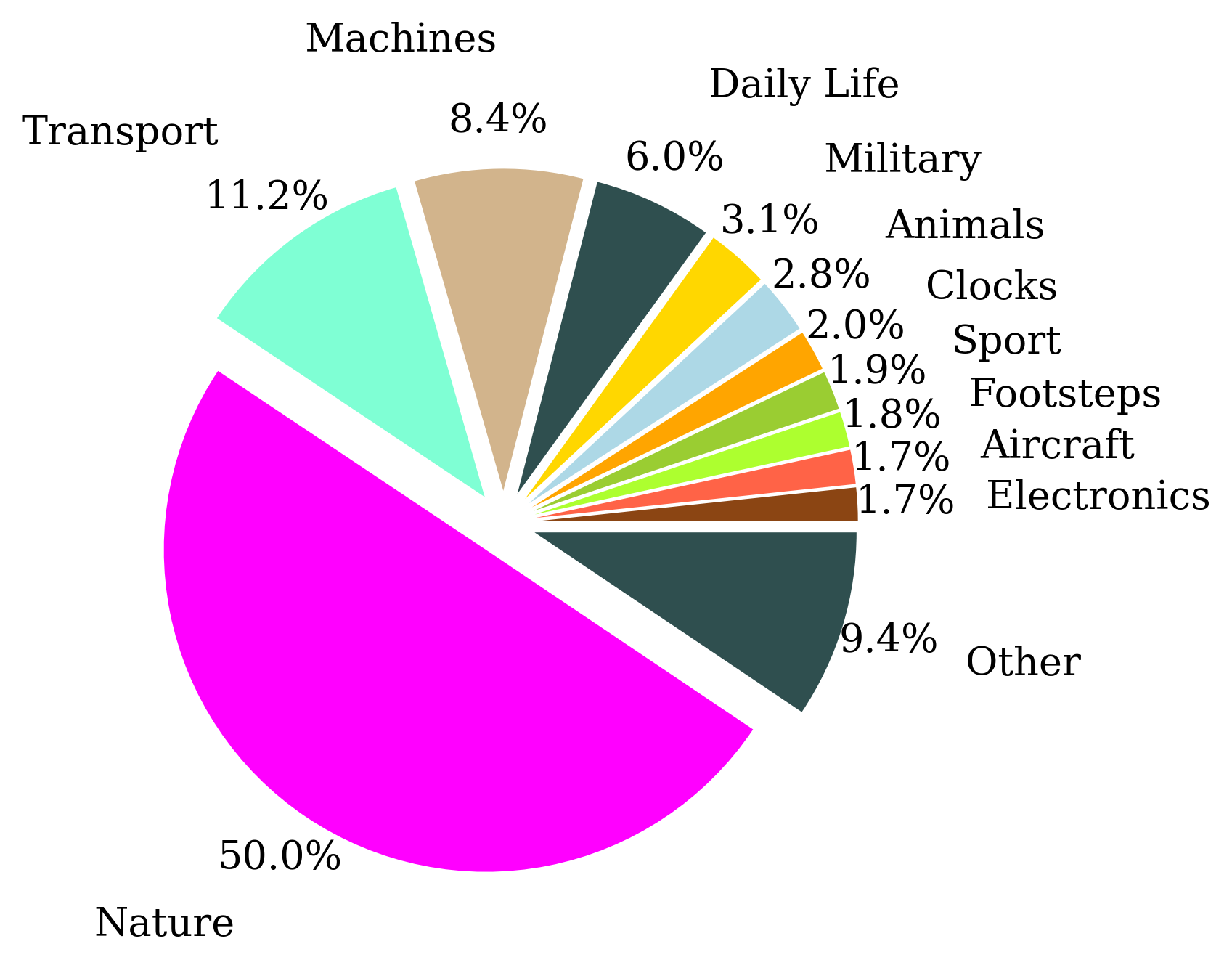}
    \caption{Pie chart showing the distribution of audio files in the \datasetName dataset over different categories.}
    \label{fig:dataset-categories}
\end{figure}

In this section, we introduce the \datasetName dataset for \task.
The \datasetName dataset consists of 32,979 audio files
accompanied by natural language descriptions.
We present an overview of the \datasetName dataset by
describing how it was collected in Section \ref{sec:data-collection},
and by providing an analysis and comparison to related datasets
which contain pairs of audio files and text descriptions in Section \ref{sec:data-analysis}.
Furthermore, we show some examples of the data in Section \ref{sec:data-examples}.

\subsection{Dataset collection}\label{sec:data-collection}
The \datasetName data was sourced from the
BBC Sound Effects webpage\footref{bbc-footnote}.
It contains audio files and corresponding textual descriptions of a wide range of sounds from the BBC Radiophonic workshop, the Blitz in London, special effects made for the BBC, and recordings from the Natural History Unit archive. The sounds are of high quality and were recorded for professional applications, such as radio and TV special effects.
In some cases, additional information is provided which contains
the size and number of channels of the audio file together with
the date it was recorded and other sound tags.
The audio files are associated with 23 categories,
including but not limited to \textit{nature}, \textit{clocks}, \textit{fire}, etc.
We show the proportion of files from the different
categories in Fig.~\ref{fig:dataset-categories}. For this, we use the text tags accompanying the collected audio files. Since some of the files contain multiple text tags, we collected all of them in a bag-of-words fashion and then provided the frequencies with which they appear.
We obtained 32,979 audio files sampled at 44.1~Hz which have a non-empty textual description
(out of the 33,066 audio files on the BBC website).
We propose to split the \datasetName dataset into
training/validation/test subsets by randomly selecting 70\%
of the files for training,
and 15\% each for validation and testing.

\subsection{Data analysis}\label{sec:data-analysis}

\begin{figure}
    \centering
    \includegraphics[width=0.43\textwidth]{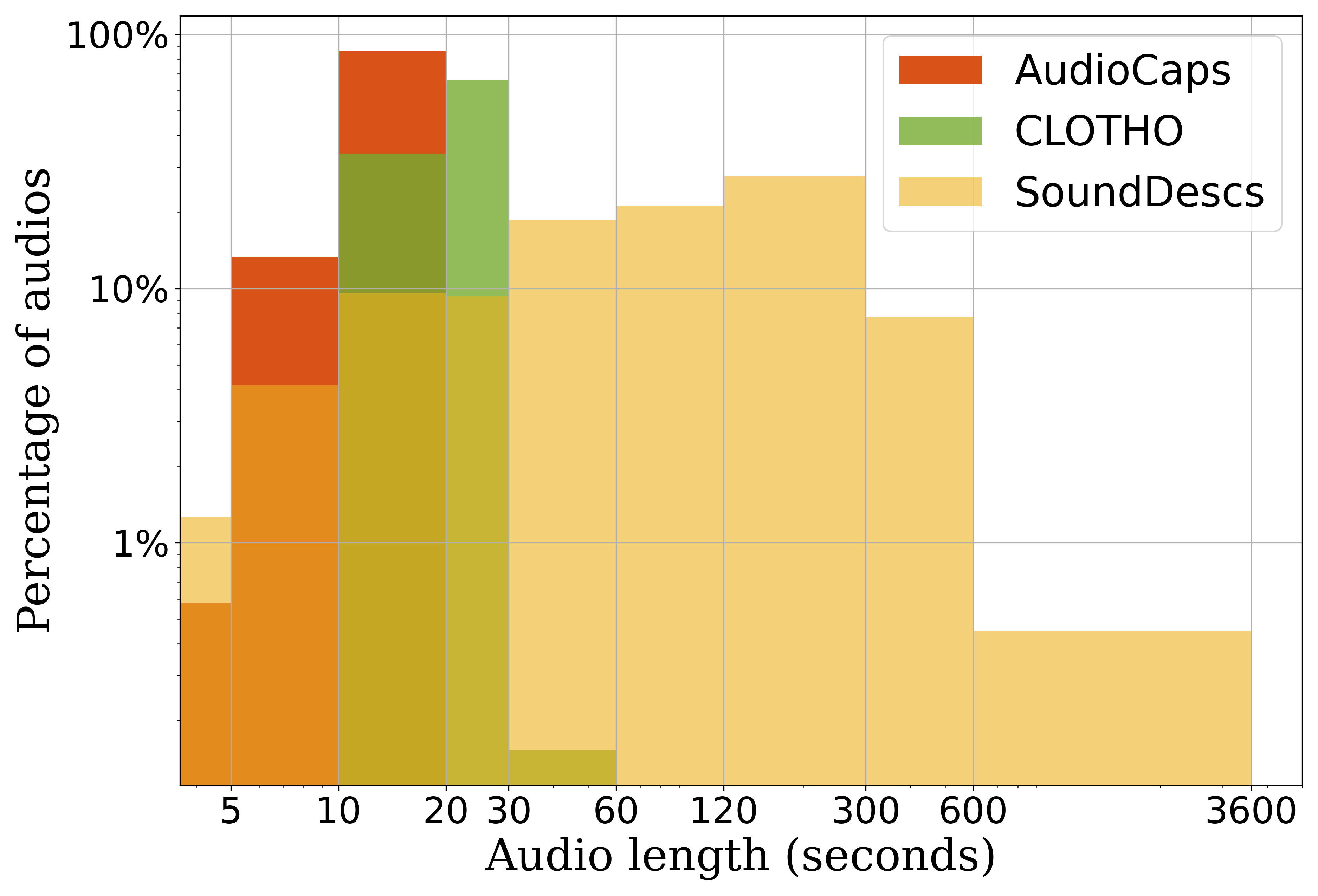}
    \caption{Histogram of audio files lengths for the \Audiocaps, \Clotho and \datasetName datasets.}
    \label{fig:audio_length}
\end{figure}

\begin{table*}[!t]
\centering
\caption{Comparative overview of sound-language datasets. Comparing the number of files in the different datasets, including their training, validation, and test subsets, audio duration (dur.) and caption lengths (\#words) measured in seconds and words respectively. Text is sourced from human caption annotations or audio descriptions provided with the sound data.
}
\setlength{\tabcolsep}{6pt}
\resizebox{\textwidth}{!}{%
\begin{tabular}{l@{\hskip 0.15cm}c c c c c | c@{\hskip 0.15cm}c@{\hskip 0.15cm}c@{\hskip 0.15cm}c | c@{\hskip 0.15cm}c@{\hskip 0.15cm}c}
\hline \hline
Dataset & Text source & language & duration(h) & \#audios &  \#captions & max dur.(s) & avg dur.(s) & max \#words & avg \#words & train & val & test\\
\hline
\Audiocaps~\cite{kim2019audiocaps} & Human captions & English & 135.01 & 50535 & 55512 & 10.08 & 9.84 & 52 & 8.80 & 49291 & 428 & 816  \\
\Clotho~\cite{drossos2020clotho} & Human captions & English & 22.55 &  3938 & 3938 & 30.00 & 22.44 & 21 & 11.32 & 2314 & 579 & 1045 \\
Audio Caption~\cite{wu2019audio}& Human captions & Chinese & 10.3 & 3707 & 3707 & N/A & 10.00 & 54 & 11.14 & 3337 &  - & 371  \\ \hline
\datasetName{} & Descriptions & English & 1060.40 & 32979 & 32979 & 4475.89 & 115.75 & 65 & 15.28 & 23085 & 4947 & 4947 \\
\hline \hline
\end{tabular}
}
\label{table:audio-datasets}
\end{table*}

\begin{figure}
    \centering
    \includegraphics[width=0.43\textwidth]{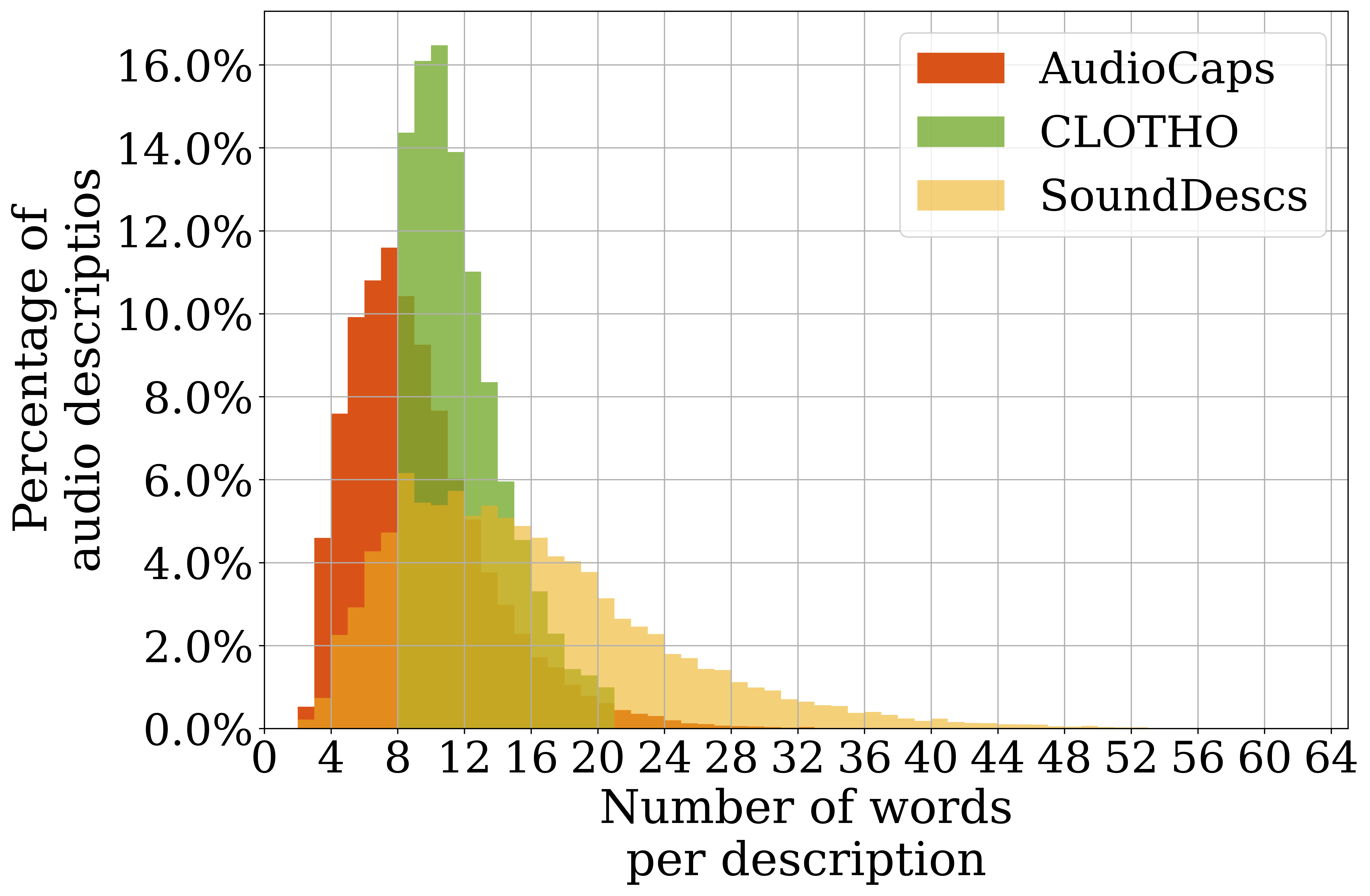}
    \caption{Distribution of the number of words per caption
    for the \Audiocaps, \Clotho and \datasetName datasets.}
    \label{fig:words_per_cap}
\vspace{-0.15cm}
\end{figure}

We compare the \datasetName dataset to related datasets which
contain matching sound and language data in Table~\ref{table:audio-datasets}.
The two main novelties of the \datasetName dataset compared to
related sound-text datasets are the wide variation in duration
of the audio files and the size of the vocabulary used for the descriptions.
The audio captioning datasets \Audiocaps~\cite{kim2019audiocaps}\footnote{The numbers provided for the \Audiocaps dataset in Table~\ref{table:audio-datasets} correspond to the subset which does not have an overlap with the VGGSound dataset.}
and \Clotho~\cite{drossos2020clotho} contain audio files that are
only 10-30 seconds long.
As can be seen in Fig. \ref{fig:audio_length}, \datasetName contains
sounds with a much wider range of audio durations with 109 files
lasting longer than 10 minutes.
Processing audio files with such variations in duration is challenging, but it enables the detailed analysis of the performance of current text-audio and audio-text retrieval methods with respect to the audio duration.
In addition to this, the \datasetName dataset is larger than all related sound-language datasets with a total duration of 1060 hours, compared to 135 hours for \Audiocaps. Since \datasetName contains fewer audio files with associated captions than \Audiocaps, the \datasetName dataset presents a more challenging retrieval benchmark dataset.
Furthermore, the average audio duration and average length of the text descriptions in \datasetName
is significantly higher than for \Audiocaps or \Clotho.
The word length distributions for these datasets can be
seen in Fig.~\ref{fig:words_per_cap}. \revv{Interestingly, for the \Clotho and \datasetName dataset which contain audio and descriptions with varied lengths, there is no strong correlation between the audio and description lengths.}

\begin{figure}
    \centering
    \includegraphics[width=0.43\textwidth]{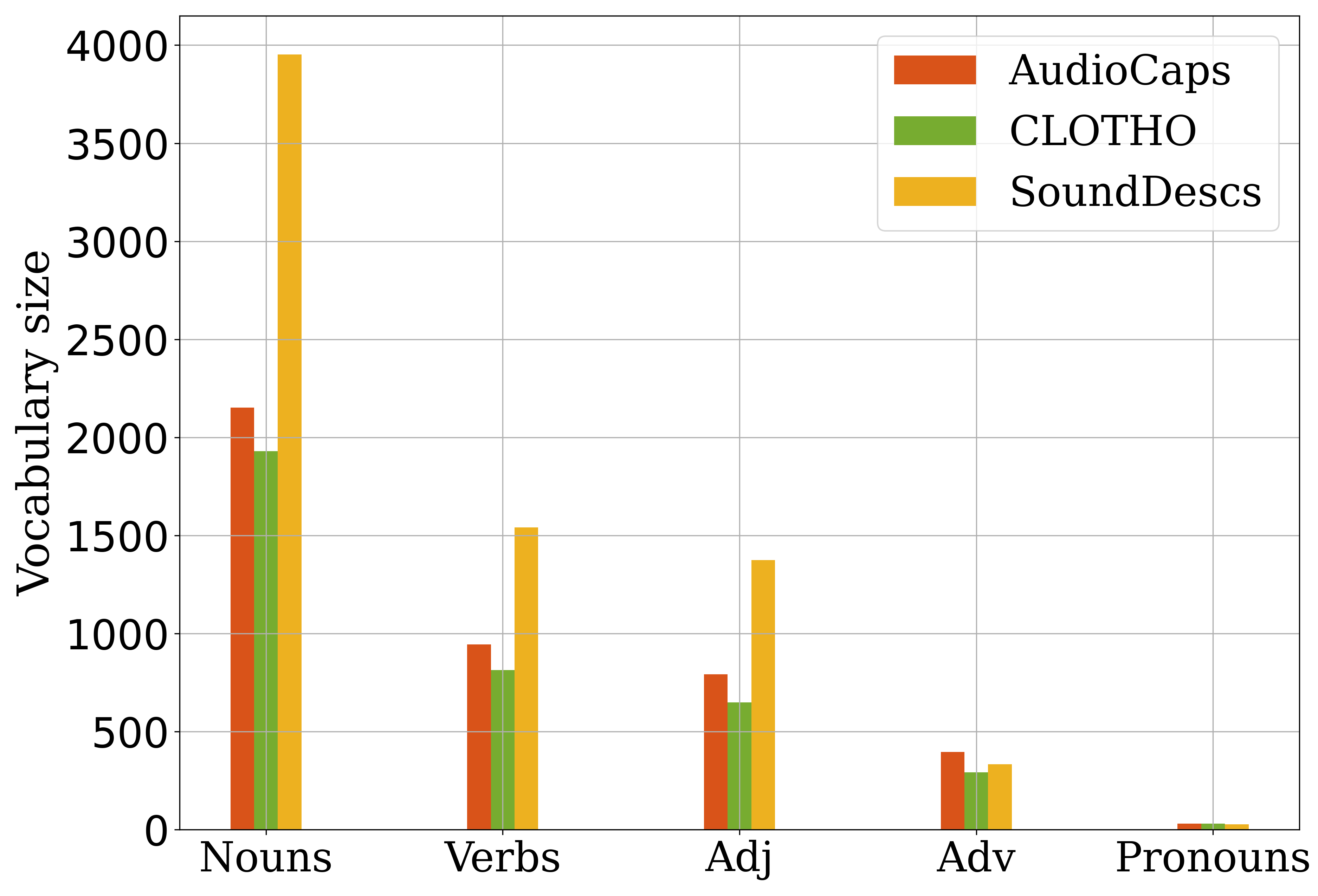}
    \caption{%
    Vocabulary size of descriptions in the \datasetName dataset,
    compared to the \Audiocaps and \Clotho datasets,
    divided into nouns, verbs, adjectives,
    adverbs, and pronouns.}
    \label{fig:pos}
\vspace{-0.2cm}
\end{figure}

Lastly, the descriptions in the \datasetName dataset have a larger vocabulary with respect to nouns, verbs, adjectives used, compared to the \Audiocaps and \Clotho datasets (see Fig.~\ref{fig:pos}). In particular, the descriptions in \datasetName contain almost 4000 distinct nouns as apposed to around 2000 in \Audiocaps and \Clotho. 
This is consistent with the wide array of topics (Fig.~\ref{fig:dataset-categories}), which reflects a high diversity of environments in which the sound effects were recorded, and thus hugely varied sources of sounds are identified in the descriptions.
A large contributor to this diversity in nouns is the high proportion of Nature sounds (Fig.~\ref{fig:dataset-categories}), for which species names are often specified (an example is shown in Fig.~\ref{fig:dataset-tsne}).

Further details about the related datasets that we use in this work can be found in Section \ref{sec:experiments}. A more in depth analysis of the \datasetName's text descriptions can be found in the Appendix.

\subsection{Dataset examples}\label{sec:data-examples}
\begin{figure}[t]
    \centering
    \includegraphics[width=0.43\textwidth]{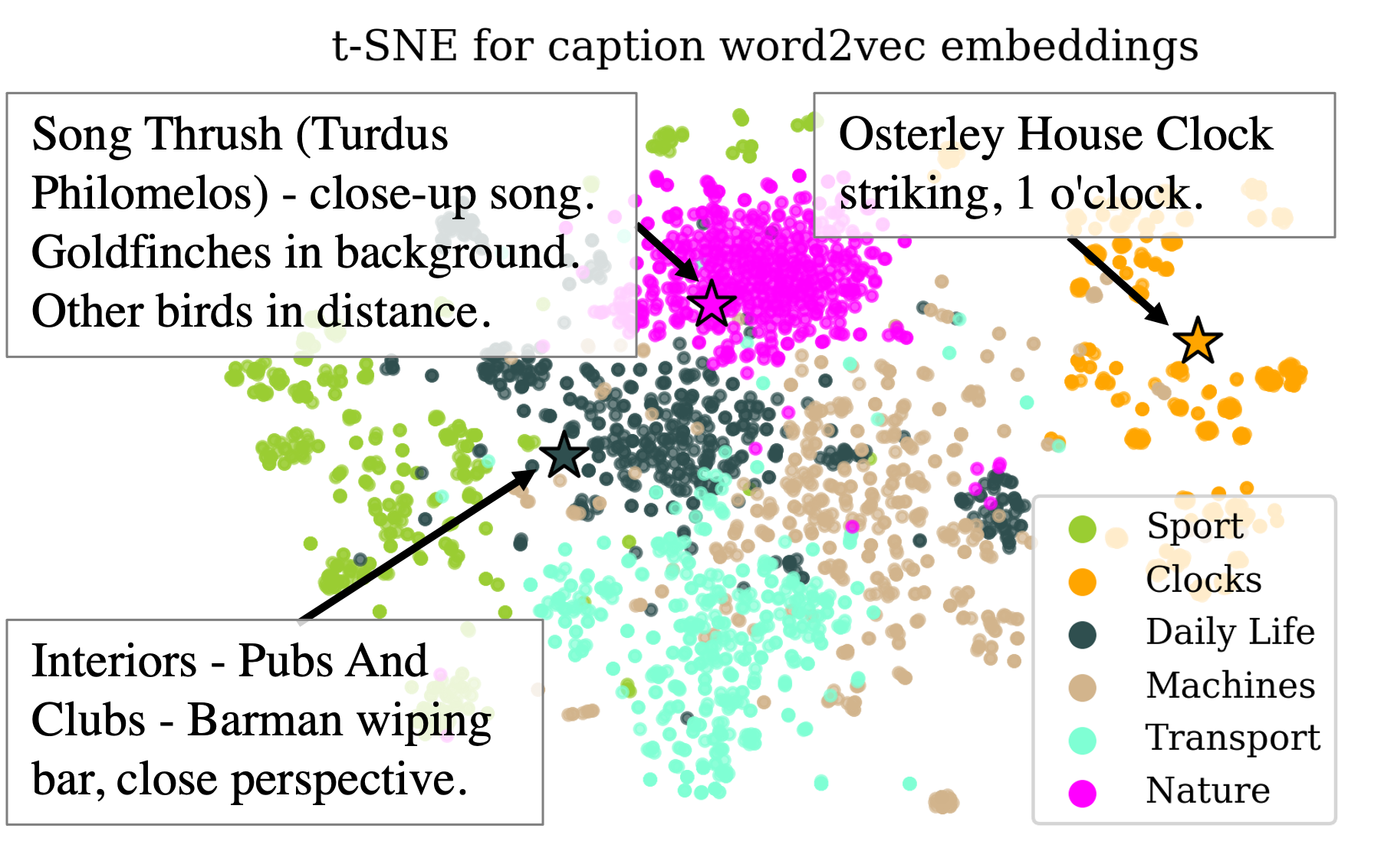}
    \caption{t-SNE visualisation for word2vec description embeddings on \datasetName. Example descriptions corresponding to embeddings marked with a star are shown in text boxes.}
    \label{fig:dataset-tsne}
\end{figure}
In Fig.~\ref{fig:dataset-tsne} we visualise the distribution of the descriptions in \datasetName, showing the full descriptions for some examples.
The averaged word2vec \cite{mikolov2013efficient} vectors extracted from each description are embedded using t-SNE \cite{van2008visualizing}, and colour-coded according to the category (Fig.~\ref{fig:dataset-categories}). We show 500 randomly chosen samples per class for 6 out of the 23 categories present in \datasetName.
We can observe that the descriptions cluster smoothly according to the categories, and that the descriptions are fairly specific and of high quality, despite not always resembling full sentences.

\subsection{Rights to use}
Data collected for generating the new \datasetName dataset
is protected by the BBC
RemArc\footnote{\url{https://sound-effects.bbcrewind.co.uk/licensing}}
licence.
This allows the data to be used for non-commercial, personal or research purposes.
We have released the list with urls along with code for downloading and constructing the
\datasetName dataset with our
proposed train/val/test split.

\section{Methods, Datasets, and Benchmark} \label{sec:method}

In this section, we first formulate the problem of audio retrieval with natural language queries.  Next, we describe three cross-modal embedding methods that we adapt for the task of audio retrieval (Section~\ref{sec:methods}). Finally, we describe the five datasets used in our experimental study (Section~\ref{sec:datasets}), and the three benchmarks that we propose for evaluating performance on the audio retrieval task (Section~\ref{sec:benchmark}).

\mypara{Problem formulation.} Given a natural language query (\textit{i.e.}~a written description of an audio event to be retrieved) and a pool of audio samples, the objective of text-audio (abbreviated to \texttt{t2a}) retrieval is to rank the audio samples according to their similarity to the query. We also consider the converse \texttt{a2t} task, viz. retrieving text with audio queries.

\subsection{Methods}\label{sec:methods}
To tackle the problem of text-audio retrieval, we propose to learn cross-modal embeddings. Specifically, given a collection of $N$ audio samples with corresponding textual descriptions, $\{(a_i, t_i) : i \in \{1, \dots, N\}\}$, we aim to learn embedding functions, $\psi_a$ and $\psi_t$, that project each audio sample $a_i$ and text sample $t_i$ into a shared space, such that $\psi_a(a_i)$ and $\psi_t(t_i)$ are close when the text describes the audio, and far apart otherwise. Writing $s_{ij}$ for the cosine similarity of the audio embedding $\psi_a(a_i)$ and the text embedding $\psi_t(t_j)$, we learn the embedding functions by minimising a contrastive ranking loss~\cite{socher2014grounded}:
\begin{align}
    \mathcal{L} = \frac{1}{B} \sum_{i=1, j \neq i}^B [m + s_{ij} - s_{ii}]_+ + [m + s_{ji} - s_{ii}]_+
    \label{eqn:loss}
\end{align}
where $B$ denotes the batch size, $m$ the \textit{margin} (set as a hyperparameter), and $[\cdot]_+ = max(\cdot, 0)$ the hinge function.

We consider three recent state-of-the-art frameworks for learning such embedding functions $\psi_a$ and $\psi_t$: Mixture-of-Embedded Experts (MoEE)~\cite{miech2018learning}, Collaborative-Experts (CE)~\cite{Liu19a}, and the Multi-modal Transformer (MMT)~\cite{gabeur2020multi}. All three frameworks were originally designed for text-video retrieval and construct their video encoder from a collection of ``experts'' (features extracted from networks pre-trained for object recognition, action classification, sound classification, etc.) which are computed for each video.

\mypara{MoEE and CE} aggregate the experts along their temporal dimension with NetVLAD~\cite{arandjelovic2016netvlad}, and project them to a lower dimension via a self-gated linear map followed by L2-normalisation. Their text encoder first embeds each word token with word2vec~\cite{mikolov2013efficient} and aggregates the results with NetVLAD~\cite{arandjelovic2016netvlad}. The result is projected by a sequence of self-gated linear maps (one for each expert) into a shared embedding space with the outputs of the video encoder. Finally, a scalar-weighted sum of the embedded experts in each joint space is used to compute the overall cosine similarity between the video and text (see~\cite{miech2018learning} for more details). CE adopts the same text encoder as MoEE and similarly makes use of multiple video experts. However, rather than projecting them directly into independent spaces against the embedded text, CE first applies a ``collaborative gating'' mechanism, which filters each expert with an element-wise attention mask that is generated with a small MLP that ingests all pairwise combinations of experts (see~\cite{Liu19a} for further details).

\mypara{MMT}, on the other hand, uses a multi-modal transformer
encoder which refines the expert embeddings by passing through
multiple multi-headed self-attention layers.
Differently from the temporal aggregation of expert features used by MoEE and CE,
expert features are instead passed as input directly to the transformer encoder.
This allows to iteratively focus on the relevant information from each expert
at multiple time steps by comparing content from different experts,
rather than down-projecting the expert embeddings in a single step
(as is the case for MoEE and CE).
Each input expert uses an aggregated embedding which collects the information
for that expert through the layers and serves as the output expert representation.
The iterative attention across different modalities and time steps
enables MMT to combine information across input experts,
rather than comparing each expert individually to the text query embeddings.
MMT leverages a pre-trained BERT model~\cite{devlin2018bert}
to extract text embeddings.
Gated embedding functions are learned to obtain a text embedding for each video expert.  
The similarity $s_{ij}$ between the aggregated audio and text embeddings
is computed as the weighted sum of the similarity between each expert
and the text embedding.
Further information about MMT can be found in \cite{gabeur2020multi}.

To adapt the MoEE, CE, and MMT frameworks for audio retrieval, we use the same
text encoder structures as used for video retrieval $\psi_t$.
We build audio encoders $\psi_a$ by mimicking the structure of
their video encoders, replacing the ``video experts'' with
``audio experts'' (described in Sec.~\ref{sec:experiments}).

\subsection{Datasets}\label{sec:datasets}
As the primary focus of our work, we study three \textit{audio-centric datasets}---these are datasets which comprise audio streams (sometimes with accompanying visual streams) paired with natural language descriptions that focus explicitly on the content of the audio track. To explore differences between audio retrieval and video retrieval, we also consider two \textit{visual-centric datasets}, that comprise audio and video streams paired with natural language which focus primarily (though not always exclusively) on the content of the video stream. Details of the five datasets we employ are given next. 

\noindent 1. \datasetName \textit{(audio-centric)} consists of sounds sourced from the BBC Sound Effects webpage and annotated with free-form text descriptions. It contains 23,085 training, 4947 validation and 4947 test samples.
\footnote{The sample list is publicly available at the project page~\cite{projectPage}.\label{project-page}} 
Further information on our newly introduced \datasetName dataset is provided in Section~\ref{sec:dataset}.

\noindent 2. \Audiocaps~\cite{kim2019audiocaps} \textit{(audio-centric)} is a dataset of sounds with event descriptions which was introduced for the task of audio captioning, with sounds sourced from the AudioSet dataset~\cite{gemmeke2017audio}.
Annotators were provided the audio tracks together with category hints (and with additional video hints if needed).
We use a subset of the data, excluding a small number of samples for which either: (i) the YouTube-hosted source video is no longer available,  (ii) the source video overlaps with the training partition of the VGGSound dataset~\cite{Chen20}.
Filtering to exclude samples affected by either issue leads to a dataset with 49,291 training, 428 validation and 816 test samples.\footref{project-page}

\noindent 3. \Clotho~\cite{drossos2020clotho} \textit{(audio-centric)} is a dataset of described sounds that was also introduced for the task of audio captioning, with sounds sourced from the Freesound platform~\cite{font2013freesound}.
During labelling, annotators only had access to the audio stream (i.e. no visual stream or meta tags) to avoid their reliance on contextual information for removing ambiguity that could not be resolved from the audio stream alone.
The descriptions are filtered to exclude transcribed speech.
The publicly available version of the dataset includes a \texttt{dev} set of 2893 audio samples and an \texttt{evaluation} set of 1045 audio samples.
Every audio sample is accompanied by 5 written descriptions.
We used a random split of the \texttt{dev} set into a training and validation set with 2,314 and 579 samples, respectively.

\noindent 4. \activityNetLong{}~\cite{krishna2017dense} (visual-centric) consists of videos sourced
from YouTube and annotated with dense event descriptions.
It allocates 10,009 videos for training and 4,917 videos for testing (we use the public \texttt{val\_1} split provided by~\cite{krishna2017dense}).
For this dataset, descriptions also tend to focus on the visual stream.

\noindent 5. \QuerYD{}~\cite{Oncescu2021} (visual-centric) is a dataset of described videos sourced from YouTube and the YouDescribe~\cite{youdescribe} platform.
It is accompanied by \textit{audio descriptions} that are provided with the explicit aim of conveying the video content to visually impaired users. Therefore, the provided descriptions focus heavily on the visual modality.
We use the version of the dataset comprising trimmed videos with 9,114 training, 1,952 validation, and 1,954 test samples.

\subsection{Benchmark}\label{sec:benchmark}
To facilitate the study of the text-based audio retrieval task, we introduce the \datasetName dataset and propose to re-purpose the two \textit{audio-centric} datasets described above, \Audiocaps and \Clotho, to provide benchmarks for text-based audio retrieval. The approach is inspired by precedents in the vision and language communities, where datasets, such as~\cite{xu2016msr}, that were originally introduced for the task of video captioning, have become popular benchmarks for text-based video retrieval~\cite{miech2018learning,wray2019fine,Liu19a,gabeur2020multi}.

\section{Experiments} \label{sec:experiments}

In this section, we first compare text-audio retrieval (\texttt{t2a})
and audio-text retrieval (\texttt{a2t}) performance on audio-centric
and visual-centric datasets.
Next, we perform an ablation study on the contributions of different
experts and present our baselines for the proposed
\Audiocaps, \Clotho, and \datasetName benchmarks.
Finally we perform experiments to assess the influence of
pre-training, audio segment duration and training dataset size,
and give qualitative examples of retrieval results.
Throughout the section, we use the standard retrieval metrics:
recall at rank $k$ (R@$k$) which measures the percentage of targets
retrieved within the top $k$ ranked results (higher is better),
along with the median (medR) and mean (meanR) rank.
For all metrics, we report the mean and standard deviation of three
different randomly seeded runs.\par

\mypara{Implementation details.}
We use pre-trained feature extractors to obtain audio and
visual expert features.
To encode the audio signal, we use two pre-trained audio
feature extractors\footnote{Since the AudioCaps test set is a
subset of the AudioSet training set (unbalanced),
we do not use audio experts pre-trained on AudioSet.}
which we refer to as \mbox{VGG}ish and \mbox{VGG}Sound.
We explain both in more detail in the following.

\mypara{VGGish.} These audio features are obtained with a \mbox{VGG}ish model \cite{Hershey16}, trained for audio classification on the YouTube-8M dataset \cite{abu2016youtube}. To produce the input for the VGGish model, the  audio stream  of  each  video  is  re-sampled  to  a  16kHz  mono  signal, converted to an STFT with a window size of 25ms and a hop size of 10ms with a Hann window, then  mapped to a 64 bin log mel spectrogram.  Finally, the features are parsed into non-overlapping 0.96s collections of frames (each collection comprises 96 frames, each of 10ms duration), which are mapped to a 128-dimensional feature vector.

\mypara{VGGSound.} These features are extracted using a ResNet-18 model \cite{He15} that has been pre-trained on the \mbox{VGG}Sound dataset (model H) \cite{Chen20}. We modify the last average pooling layer to aggregate along the frequency dimension, but keep the full temporal dimension. This results in features of dimension t x 512, where t denotes the number of time steps.

For the results with visual experts in Table \ref{table:full-ce-audiocaps}, we employed a subset of visual feature extractors used in \cite{Liu19a} which we refer to as Inst, Scene, and R2P1D. We we will describe each of those in more detail below.

\mypara{Inst.}
These features are extracted using a ResNeXt-101 model \cite{xu2015jointly} that has been pre-trained on Instagram hashtags \cite{mahajan2018exploring} and finetuned on ImageNet \cite{Deng09} for the task of image classification. Features are extracted from frames extracted at 25 fps, where each frame is resized to 224 × 224 pixels. Embeddings are 2048-dimensional.

\mypara{Scene.}
Scene features are extracted from 224×224 pixel centre crops with a DenseNet-161 model \cite{huang2017densely} pre-trained on Places365 \cite{zhou2017places}. Embeddings are 2208-dimensional.

\mypara{R2P1D.}
Features are extracted with a 34-layer R(2+1)D model \cite{tran2018closer} trained on IG-65M \cite{ghadiyaram2019large} which processes clips of 8 consecutive 112 × 112 pixel frames, extracted at 30 fps. Embeddings are 512-dimensional.

\begin{table*}[!ht]
\centering
\caption{Audio retrieval on \Audiocaps{}, \Clotho{}, and \datasetName{}, using the CE, MoEE, and MMT frameworks. Retrieval performance metrics for text to audio and audio to text retrieval reported are recall at rank k (R@1, R@5, R@10), and the median and mean rank (medR and meanR). Audio experts used are obtained from the VGGish model~\cite{gemmeke2017audio} and from a ResNet18 model pre-trained on VGGSound~\cite{Chen20}. 
}
\resizebox{2\columnwidth}{!}{%
\begin{tabular}{l@{\hskip 0.15cm}|cccccc|cccccc}
\hline \hline
\multicolumn{1}{c}{} &
\multicolumn{6}{c}{\hspace{-0.2cm}Text $\implies$ Audio} & \multicolumn{6}{c}{\hspace{-0.2cm}Audio $\implies$ Text} \\
Dataset & R$@$1 $\uparrow$ & R$@$5 $\uparrow$ & R$@$10 $\uparrow$  & R$@$50 $\uparrow$ & medR $\downarrow$ & meanR $\downarrow$ & R$@$1 $\uparrow$ & R$@$5 $\uparrow$ & R$@$10 $\uparrow$  & R$@$50 $\uparrow$ & medR $\downarrow$ & meanR $\downarrow$  \\
\hline
{\textbf{\textit{\Audiocaps{}}}} &&&&&&&&&&&\\
CE   & $\rev{23.6_{\pm0.6}}$ & $\rev{56.2_{\pm0.5}}$ & $\rev{71.4_{\pm0.5}}$ & $\rev{92.3_{\pm1.5}}$ & $\rev{4.0_{\pm0.0}}$ & $\rev{18.3_{\pm3.0}}$ & $\rev{27.6_{\pm1.0}}$ & $\rev{60.5_{\pm0.7}}$ &	$\rev{74.7_{\pm0.8}}$ & $\rev{94.2_{\pm0.4}}$ &	$\rev{4.0_{\pm0.0}}$ & $\rev{14.7_{\pm1.4}}$\\
MoEE   & $\rev{23.0_{\pm0.7}}$& $\rev{55.7_{\pm0.3}}$& $\rev{71.0_{\pm1.2}}$& $\rev{93.0_{\pm0.3}}$& $\rev{4.0_{\pm0.0}}$& $\rev{16.3_{\pm0.5}}$& $\rev{26.6_{\pm0.7}}$& $\rev{59.3_{\pm1.4}}$& $\rev{73.5_{\pm1.1}}$& $\rev{94.0_{\pm0.5}}$& $\rev{4.0_{\pm0.0}}$& $\rev{15.6_{\pm0.8}}$\\
MMT & $\rev{36.1_{\pm3.3}}$& $\rev{72.0_{\pm2.9}}$& $\rev{84.5_{\pm2.0}}$& $\rev{97.6_{\pm0.4}}$& $\rev{2.3_{\pm0.6}}$& $\rev{7.5_{\pm1.3}}$& $\rev{39.6_{\pm0.2}}$& $\rev{76.8_{\pm0.9}}$& $\rev{86.7_{\pm1.8}}$& $\rev{98.2_{\pm0.4}}$& $\rev{2.0_{\pm0.0}}$& $\rev{6.5_{\pm0.5}}$ \\ \hline 
{\textbf{\textit{\Clotho{}}}} &&&&&&&&&&&\\
CE   & $6.7_{\pm0.4}$ & $	21.6_{\pm0.6}$ & $	33.2_{\pm0.3}$ & $	69.8_{\pm0.3}$ & $	22.3_{\pm0.6}$ & $	58.3_{\pm1.1}$ & $ 7.0_{\pm0.3}$ & $	22.7_{\pm0.6}$ & $	34.6_{\pm0.5}$ & $	67.9_{\pm2.3}$ & $	21.3_{\pm0.6}$ & $	72.6_{\pm3.4}$\\
MoEE   & $6.0_{\pm0.1}$& $20.8_{\pm0.7}$& $32.3_{\pm0.3}$& $68.5_{\pm0.5}$& $23.0_{\pm0.0}$& $60.2_{\pm0.8}$& $7.2_{\pm0.5}$& $22.1_{\pm0.7}$& $33.2_{\pm1.1}$& $67.4_{\pm0.3}$& $22.7_{\pm0.6}$& $71.8_{\pm2.3}$\\
MMT & $6.5_{\pm0.6}$& $21.6_{\pm0.7}$& $32.8_{\pm2.1}$& $66.9_{\pm2.0}$& $23.0_{\pm2.6}$& $67.7_{\pm3.1}$& $6.3_{\pm0.5}$& $22.8_{\pm1.7}$& $33.3_{\pm2.2}$& $67.8_{\pm1.5}$& $22.3_{\pm1.5}$& $67.3_{\pm2.9}$ \\ \hline

{\textbf{\textit{\datasetName{}}}} &&&&&&&&&&&\\
CE   & $31.1_{\pm0.2}$	& $60.6_{\pm0.7}$ &	$70.8_{\pm0.5}$ & $86.0_{\pm0.2}$ & $3.0_{\pm0.0}$ & $63.6_{\pm2.2}$ & $30.8_{\pm0.8}$ & $60.3_{\pm0.3}$ & $69.5_{\pm0.1}$ & $85.4_{\pm0.2}$ & $3.0_{\pm0.0}$ & $63.2_{\pm0.6}$\\
MoEE   & $30.8_{\pm0.7}$	& $60.8_{\pm0.3}$ &	$70.9_{\pm0.5}$ & $85.9_{\pm0.6}$ & $3.0_{\pm0.0}$ & $62.0_{\pm3.8}$ & $30.9_{\pm0.3}$ & $60.3_{\pm0.4}$ & $70.1_{\pm0.3}$ & $85.3_{\pm0.6}$ & $3.0_{\pm0.0}$ & $61.5_{\pm3.2}$\\
MMT   & $30.7_{\pm0.4}$& $61.8_{\pm1.0}$& $72.2_{\pm0.8}$& $88.8_{\pm0.4}$& $3.0_{\pm0.0}$& $34.0_{\pm0.6}$& $31.4_{\pm0.8}$& $63.2_{\pm0.7}$& $73.4_{\pm0.5}$& $89.0_{\pm0.3}$& $3.0_{\pm0.0}$& $32.5_{\pm0.4}$\\

\hline \hline
\end{tabular}
}
\label{table:mmt_ce_all}
\end{table*}

\mypara{Training}
All models were trained using the contrastive ranking loss (Eqn.~\ref{eqn:loss}), with $m$ set to 0.2 for CE and MoEE, and 0.05 for MMT (those parameters were taken from \cite{Liu19a} and \cite{gabeur2020multi} respectively). CE and MoEE models were trained with a batch size of 128 for 20 epochs and the models that gave the best performance on the geometric mean of R@1, R@5, and R@10 were chosen as the final models. MMT was trained with a batch size of 32 for 50K steps.

For CE and MoEE, we used the Lookahead solver \cite{zhang2019lookahead} in combination with RAdam \cite{liu2019variance} (implementation by \cite{lesswright}) with an initial learning rate of $0.01$ and weight decay of $0.001$. We use a learning rate decay for each parameter group with a factor of $0.95$ every epoch.
MMT was trained using Adam and a learning rate of $0.00005$, which was decayed by a multiplicative factor 0.95 every 1K optimisation steps.

For the NetVLAD module in CE and MoEE,
we used 20 VLAD clusters and one ghost cluster~\cite{zhong2018ghostvlad}
for text, and 16 VLAD clusters for the audio features. 
On \Audiocaps, we used a maximum of 52 word tokens,
a maximum of 10 time frames for VGGish and a maximum of 32 time frames for VGGSound features.
For \Clotho, we used a maximum of 21 word tokens,
a maximum of 31 time frames for VGGish,
and 95 VGGSound features per sample
(95 for both audio feature experts for MMT).
For \datasetName, the maximum number of word
tokens was set to 46, and audio time frames used to 400 (both for VGGish and VGGSound).

\begin{table}[t]
\centering
\caption{Audio retrieval on audio-centric and visual-centric datasets. Performance is strongest on the audio-centric \datasetName{} dataset and weakest on the visual-centric \activityNetLong{} (\activityNetShort) dataset.
}
\footnotesize
\setlength{\tabcolsep}{6pt}
\resizebox{\columnwidth}{!}{%
\begin{tabular}{l@{\hskip 0.15cm}c@{\hskip 0.2cm}c | @{\hskip 0.1cm}c@{\hskip 0.1cm}c|@{\hskip 0.1cm}c @{\hskip 0.1cm}c}
\hline \hline
\multicolumn{3}{c}{} &
\multicolumn{2}{c}{\hspace{-0.2cm}Text $\implies$ Audio} & \multicolumn{2}{c}{\hspace{-0.2cm}Audio $\implies$ Text} \\
Dataset & Anno. Focus & Pool & R$@$1 $\uparrow$ & R$@$10 $\uparrow$  & R$@$1 $\uparrow$ & R$@$10 $\uparrow$\\
\hline
\Audiocaps~\cite{kim2019audiocaps} & audio & 816 & 
$\rev{18.5_{\pm0.3}}$&$\rev{62.0_{\pm0.5}}$&$ \rev{20.7_{\pm1.8}} $&$\rev{62.9_{\pm0.4}}$ \\
\Clotho~\cite{drossos2020clotho} & audio & 1045 & $4.0_{\pm0.2}$&$25.4_{\pm0.5}$&$ 4.8_{\pm0.4} $&$25.8_{\pm1.7}$\\
\datasetName{} & audio & 4947 & $25.4_{\pm0.6}$ & $64.1_{\pm0.3}$ & $24.2_{\pm0.3}$ & $62.5_{\pm0.2}$\\
\hline
\activityNetShort{}~\cite{krishna2017dense} & visual & 4917 &$1.4_{\pm0.1}$&$8.5_{\pm0.2}$&$ 1.1_{\pm0.1} $&$7.9_{\pm0.0}$ \\
\QuerYD~\cite{Oncescu2021} & visual & 1954 & $3.7_{\pm0.2}$&$17.3_{\pm0.6}$&$ 3.8_{\pm0.2} $&$16.8_{\pm0.2}$ \\

\hline \hline
\end{tabular}
}

\label{table:audio-retrieval-datasets}
\end{table}

\begin{table}
\centering
\caption{The influence of different experts on \Audiocaps{}. A comparison of audio and visual experts (applied to the video from which the audio was sourced) using CE~\cite{Liu19a}. Audio features are significantly more effective than visual features (which nevertheless provide some complementary signal as can be seen when jointly using audio and visual features).
}
\footnotesize
\setlength{\tabcolsep}{6pt}
\resizebox{\columnwidth}{!}{%
\begin{tabular}{l | @{\hskip 0.1cm}c@{\hskip 0.1cm}c|@{\hskip 0.1cm}c @{\hskip 0.1cm}c}
\hline \hline
\multicolumn{1}{c}{} &
\multicolumn{2}{c}{\hspace{-0.2cm}Text $\implies$ Audio/Video} & \multicolumn{2}{c}{\hspace{-0.2cm}Audio/Video $\implies$ Text} \\
Expert & R$@$1  $\uparrow$ & R$@$10 $\uparrow$  & R$@$1 $\uparrow$ & R$@$10 $\uparrow$\\
\hline
\textbf{Visual experts only} &&&&\\
Scene & $\rev{6.0_{\pm0.0}}$&$\rev{35.6_{\pm0.8}}$& $\rev{6.8_{\pm0.6}}$&$\rev{31.9_{\pm1.3}}$ \\
Inst & $\rev{8.2_{\pm0.3}}$&$\rev{46.2_{\pm0.5}}$&$\rev{10.1_{\pm0.8}} $&$\rev{41.3_{\pm0.6}}$\\
R2P1D & $\rev{8.1_{\pm0.4}}$&$\rev{45.8_{\pm0.2}}$&$ \rev{10.7_{\pm0.1}} $&$\rev{43.4_{\pm1.9}}$\\
Scene + Inst & $\rev{8.2_{\pm0.3}}$&$\rev{47.1_{\pm0.2}}$& $\rev{10.2_{\pm1.2}} $&$\rev{41.5_{\pm1.3}}$ \\
Scene + R2P1D & $\rev{8.6_{\pm0.1}}$&$\rev{47.4_{\pm0.2}}$&$ \rev{11.6_{\pm0.4}} $&$\rev{43.5_{\pm0.8}}$ \\
R2P1D + Inst (\textit{CE-Visual}) & $ \rev{\bm{9.5_{\pm0.6}}}$&$\rev{\bm{50.0_{\pm0.5}}}$&$ \rev{\bm{11.2_{\pm0.1}}} $&$\rev{\bm{45.2_{\pm1.9}}}$ \\
\hline
\textbf{Audio experts only} &&&&\\
VGGish & $\rev{18.5_{\pm0.3}}$&$\rev{62.0_{\pm0.5}}$&$ \rev{20.7_{\pm1.8}} $&$\rev{62.9_{\pm0.4}}$ \\
VGGSound & $\rev{22.4_{\pm0.3}}$&$\rev{69.2_{\pm0.9}}$&$ \rev{27.0_{\pm0.9}} $&$\rev{72.5_{\pm0.7}}$\\
VGGish + VGGSound (\textit{CE-Audio}) & $\rev{\bm{23.6_{\pm0.6}}}$&$\rev{\bm{71.4_{\pm0.5}}}$&$ \rev{\bm{27.6_{\pm1.0}}} $&$\rev{\bm{74.7_{\pm0.8}}}$ \\
\hline
\textbf{Audio and visual experts} &&&&\\
\textit{CE-Visual} + VGGish & $\rev{24.5_{\pm0.8}}$&$\rev{74.9_{\pm1.0}}$&$ \rev{31.0_{\pm2.2}} $&$\rev{78.8_{\pm1.2}}$\\
\textit{CE-Visual} + VGGSound & $\rev{27.6_{\pm0.2}}$&$\rev{78.0_{\pm0.8}}$&$ \rev{32.7_{\pm0.9}} $&$\rev{82.4_{\pm0.4}}$ \\
\textit{CE-Visual} + \textit{CE-Audio}  & $\rev{\bm{28.0_{\pm0.5}}}$&$\rev{\bm{80.4_{\pm0.3}}}$&$ \rev{\bm{35.8_{\pm0.6}}} $&$\rev{\bm{83.3_{\pm0.6}}}$\\
\hline \hline
\end{tabular}
}
\label{table:full-ce-audiocaps}
\end{table}

\begin{figure*}[t]
\centering
\raisebox{0cm}{\href{https://www.robots.ox.ac.uk/~vgg/research/audio-retrieval/}{\includegraphics[width=0.99\textwidth,clip,trim={0cm 0cm 0cm 0cm}]{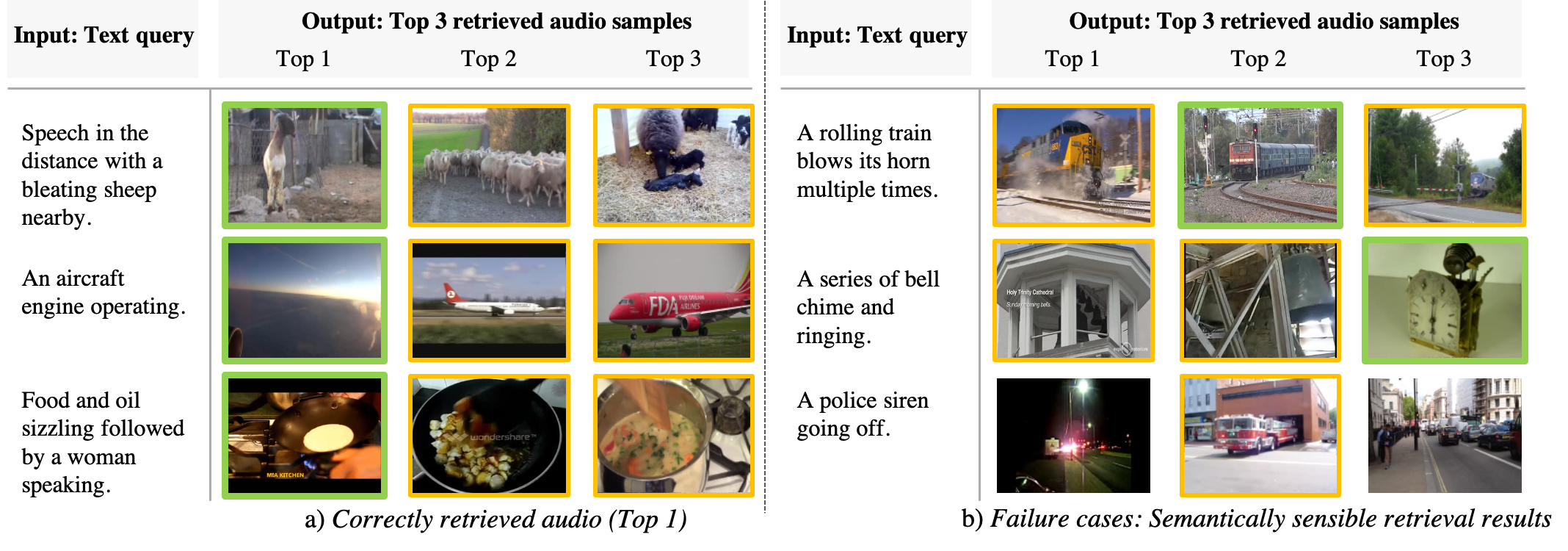}}}
\caption{Qualitative results. Text-based audio retrieval results on \Audiocaps using CE with VGGish and VGGSound features. For an input text query, we visualise the top 3 retrieved audio samples using a video frame from the corresponding videos (the audio can be heard at the project webpage~\cite{projectPage}). We mark the audio samples which correspond to the query with green boxes. 
Successful retrievals are shown in a), failures in b). Note, in particular, the examples in b), where the model's top 1 retrieved audio is not the correct one, but the retrieved results nevertheless sound reasonable (visually convincing results are marked with yellow boxes).
}
\label{fig:qualitative}
\end{figure*}

\mypara{Audio-centric vs. visual-centric queries.} 
We first investigate audio retrieval with audio-centric
and visual-centric queries (The audio-centric datasets contain audio-centric queries, and the video-centric datasets contain video-centric queries.).
For this experiment, we use CE with a single
expert (\mbox{VGG}ish audio features).
In Table~\ref{table:audio-retrieval-datasets},
we observe that performance is strongest on the
audio-centric datasets \datasetName and \Audiocaps,
and it is weakest overall on the visual-centric \activityNetLong dataset.
This is expected, since visual-centric queries contain information that is not captured in the audio data.
We note that the \Clotho dataset is particularly challenging, with performance weaker (accounting for pool size) than the visual-centric \QuerYD{} dataset. We hypothesise that this is for two reasons: (1) the significantly smaller training set size of \Clotho, compared to all other datasets, (2) \Clotho was constructed such that the audio tag distribution resulted in varied audio content,
making it a potentially more difficult benchmark.
We note, however, that the \QuerYD{} experiments suggest that computationally efficient video retrieval using only the audio stream can still be obtained, although at a lower accuracy.\par

\mypara{Ablation study.}
We next conduct an ablation study to investigate the effectiveness of different audio and visual experts for audio retrieval on the \Audiocaps dataset. We perform this experiment on the \Audiocaps dataset, since the \datasetName and \Clotho datasets are audio-only datasets which implies that we cannot use any visual experts.
We present the results in Table~\ref{table:full-ce-audiocaps}, where we observe that audio experts significantly outperform visual experts (pre-trained models for visual tasks like scene classification, which we compute from the video from which the audio was sourced). We note that the combination of audio and visual experts performs strongest overall, suggesting that the audio-centric queries contain information that is more accessible from the visual modality. The strongest audio-only retrieval is achieved by combining \mbox{VGG}ish and \mbox{VGG}Sound features---we therefore adopt this setting for the remaining experiments.

Additionally, we experimented with using speech features (word2vec~\cite{mikolov2013efficient} encodings of speech-to-text transcriptions~\cite{google-speech} of the audio stream). However, this did not improve the retrieval performance. Upon further investigation, we found that the audio captions and the spoken words in the \Audiocaps dataset do not have a significant overlap (corresponding to a METEOR~\cite{banerjee2005meteor} score of only 0.03 -- perfect agreement between text sentences would give a score of 1).

\begin{table}
\centering
\caption{Pre-training for audio retrieval. Text-audio and audio-text retrieval results for CE~\cite{Liu19a} with VGGish and VGGSound features on the proposed \datasetName{}, \Audiocaps{}, and \Clotho{} retrieval benchmarks. Pre-training on \Audiocaps{} improves the performance on \Clotho{}, and pre-training on \datasetName{} slightly boosts the performance on \Audiocaps{}.
}
\footnotesize
\setlength{\tabcolsep}{6pt}
\begin{tabular}{l |@{\hskip 0.2cm}c  @{\hskip 0.1cm}c | @{\hskip 0.1cm}c @{\hskip 0.1cm}c }
\hline \hline
 &
\multicolumn{2}{c}{Text $\implies$ Audio} & \multicolumn{2}{c}{Audio $\implies$ Text} \\
Pre-training & R$@$1 $\uparrow$ & R$@$10 $\uparrow$  & R$@$1 $\uparrow$ & R$@$10 $\uparrow$\\
\hline
\textbf{\Audiocaps{}} &&&\\
None & $\rev{23.6_{\pm0.6}}$&$\rev{71.4_{\pm0.5}}$&$ \rev{27.6_{\pm1.0}} $&$\rev{74.7_{\pm0.8}}$ \\
\datasetName{}  & $\rev{24.6_{\pm0.1}}$&$\rev{72.2_{\pm0.8}}$&$ \rev{27.8_{\pm0.6}} $&$\rev{75.2_{\pm0.4}}$\\
\hline 
\textbf{\Clotho{}} &&&&\\
None &$6.7_{\pm0.4}$&$33.2_{\pm0.3}$&$ 7.0_{\pm0.3} $&$34.6_{\pm0.5}$\\
\Audiocaps{}& $\rev{9.1_{\pm0.3}}$&$\rev{39.7_{\pm0.4}}$&$ \rev{11.1_{\pm1.1}} $&$\rev{39.6_{\pm1.1}}$\\
\datasetName{} & $6.4_{\pm0.5}$&$32.5_{\pm1.7}$&$ 6.1_{\pm0.7} $&$31.4_{\pm1.8}$\\
\hline 
\textbf{\datasetName{}} &&&&\\
None & $31.1_{\pm0.2}$&$70.8_{\pm0.5}$&$ 30.8_{\pm0.8} $&$69.5_{\pm0.1}$\\
\Audiocaps{} & $\rev{23.3_{\pm0.7}}$&$\rev{63.9_{\pm0.5}}$&$ \rev{22.2_{\pm0.4}} $&$\rev{63.3_{\pm0.3}}$\\
\hline \hline
\end{tabular}

\label{table:finetuning}
\end{table}

\mypara{Benchmark results.} Incorporating the strongest combination of experts from the ablation study, we report our final baselines for text-audio and audio-text retrieval for three methods on the \datasetName, \Audiocaps, and \Clotho datasets in Table~\ref{table:mmt_ce_all}. We observe that MMT outperforms CE and MoEE on the \Audiocaps and \Clotho datasets for both text-audio and audio-text retrieval. For \datasetName, all three models yield comparable results, with CE and MoEE being slightly stronger than MMT.

We also report the performance after pre-training the CE model for retrieval on the \Audiocaps or \datasetName datasets and then fine-tuning on \Clotho, \Audiocaps, or \datasetName in Table \ref{table:finetuning}. Here, we observe that pre-training on the \datasetName brings a slight boost for \Audiocaps and harms the performance on \Clotho. This might be due to a larger domain gap between \datasetName and \Clotho compared to \Audiocaps. Pre-training on \Audiocaps improves the performance on \Clotho, but is not beneficial for \datasetName. 
Furthermore, we explored pre-training on \Clotho and fine-tuning on the \Audiocaps and \datasetName datasets, but found negligible change in performance (likely due to the fact that the \Audiocaps training set is significantly larger than that of \Clotho).

\mypara{Qualitative results.}  The qualitative results in Fig. \ref{fig:qualitative} show examples in which the CE model with \mbox{VGG}ish and \mbox{VGG}Sound expert modalities (CE-Audio) is used to retrieve audio with natural language queries. The retrieved results mostly contain audio that is semantically similar to the input text queries. Observed failure cases arise from audio samples sounding very similar to one another despite being semantically distinct (e.g. the siren of a fire engine sounds very similar to a police siren).

\begin{figure}[t]
\centering
\begin{subfigure}[b]{0.215\textwidth}
\centering
\includegraphics[width=1\textwidth,clip,trim={0.2cm 0cm 0cm 0cm}]{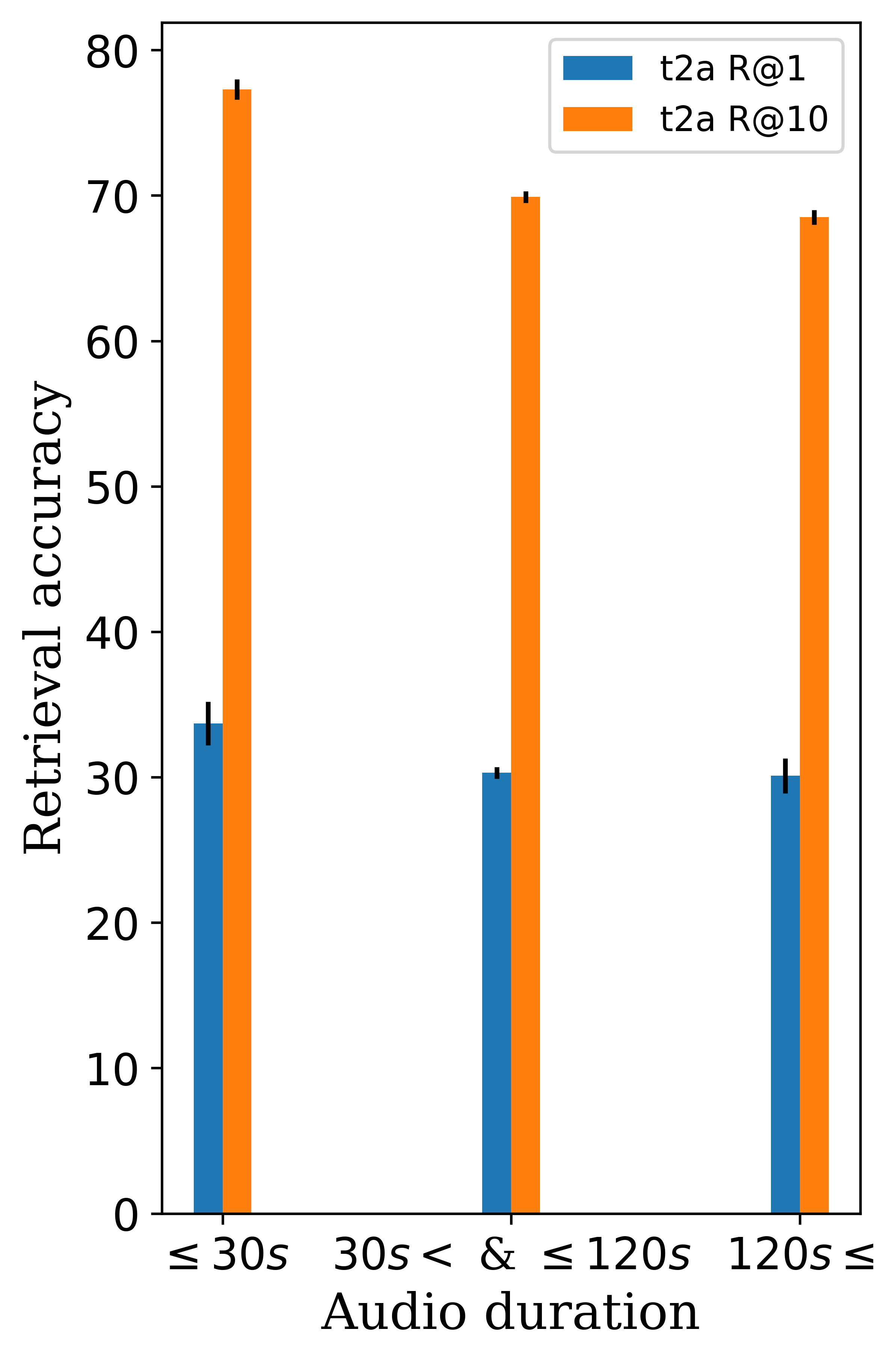}
\caption{Audio duration.}\label{fig:duration-sounddescs}
\end{subfigure}
\begin{subfigure}[b]{0.22\textwidth}
\centering
\includegraphics[width=1\textwidth,clip,trim={0.2cm 0cm 0cm 0cm}]{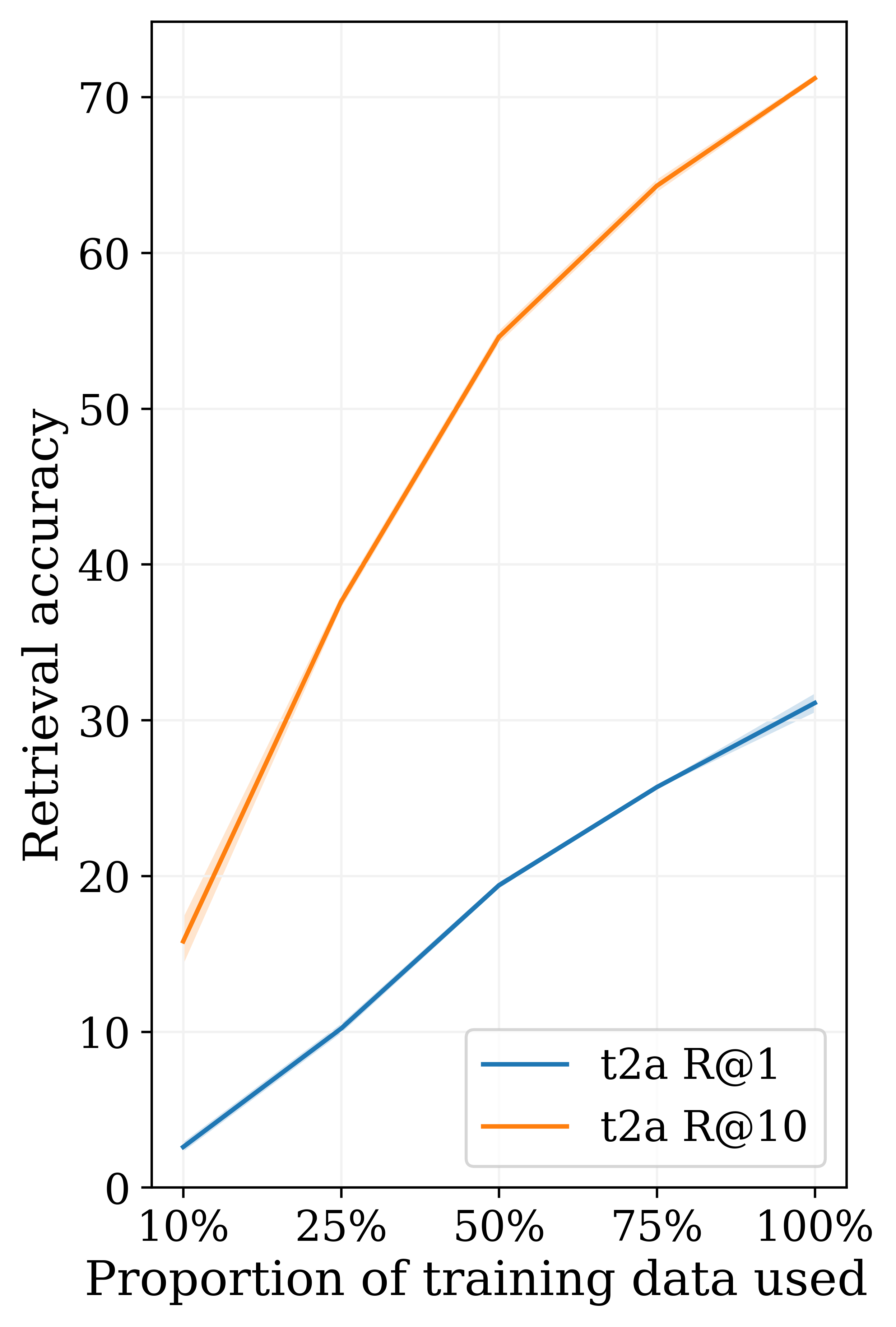}
\caption{Training data size.}\label{fig:scale-bbcsound}
\end{subfigure}
\caption{The influence of audio duration and training data scale on the audio retrieval performance on \datasetName. Performance for the CE model is shown for the subsets of the \datasetName test set according to different audio durations in a), and for different proportions of available training data in b).}
\label{fig:trainset_ablation}
\end{figure}

\mypara{Influence of audio segment duration.}
Next, we investigate the influence of audio segment duration on
retrieval accuracy on the \datasetName dataset. The retrieval accuracy for CE on different subsets of the test data according to their audio duration is shown in Fig.~\ref{fig:duration-sounddescs}. We show the performance for \datasetName audio files with a duration up to 30 seconds, those between 30 and 120 seconds, and those that are longer than 120 seconds.
We observe that the retrieval performance is slightly weaker for longer audio segments than for those that last less than 30 seconds. However, the performance for very long audio files (longer than 120 seconds) is still solid. 

\mypara{Influence of training scale.}
Finally, we present experiments using different proportions of the \datasetName dataset for training CE-Audio in Fig.~\ref{fig:scale-bbcsound}.
As expected for deep learning frameworks, we observe that as more training data becomes available, the performance increases monotonically.
We also observe that there is still clear room for improvement
in terms of retrieval results simply by collecting additional
training data, motivating further dataset construction work to support
future research on this important task.

\section{Conclusion} \label{sec:conclusion}
We introduced the novel \datasetName dataset for natural language based audio retrieval. Furthermore, we proposed three benchmarks for natural language based audio retrieval on the \Clotho, \Audiocaps, and \datasetName datasets, and provided baseline results by adapting strong multi-modal video retrieval methods.
Our results show that these methods are relatively well-suited for the audio retrieval task, however there is room for improvement, as expected for an under-explored problem.
We hope that our proposed benchmarks will facilitate the development of future audio search engines, and make this large fraction of the world's produced media available for public use.

\ifCLASSOPTIONcompsoc
  \section*{Acknowledgments}
\else
  \section*{Acknowledgment}
\fi

ASK and ZA were supported by the ERC (853489 - DEXIM), by the DFG (2064/1 – Project number 390727645), and by the BMBF (FKZ: 01IS18039A). AMO was supported by an EPSRC DTA Studentship.
JFH is supported by the Royal Academy of Engineering
(RF\textbackslash201819\textbackslash18\textbackslash163).
SA was supported by EPSRC EP/T028572/1 Visual AI.
The authors would like to thank A. Zisserman
for suggestions.
SA would also like to thank Z. Novak and
S. Carlson for support.

\appendix
\begin{section}{Supplementary Material}\label{sec:supplementary}
In this appendix, we provide additional information about the \datasetName dataset. \par

In Fig.~\ref{fig:words_per_cap}, we presented a comparison of the description length distributions for the \Audiocaps, \Clotho, and \datasetName datasets. Since descriptions in \datasetName can contain one or more sentences, we also provide the distribution of sentence lengths in Fig.~\ref{fig:words_per_sent}.

\begin{figure}[h]
    \centering
    \includegraphics[width=0.30\textwidth]{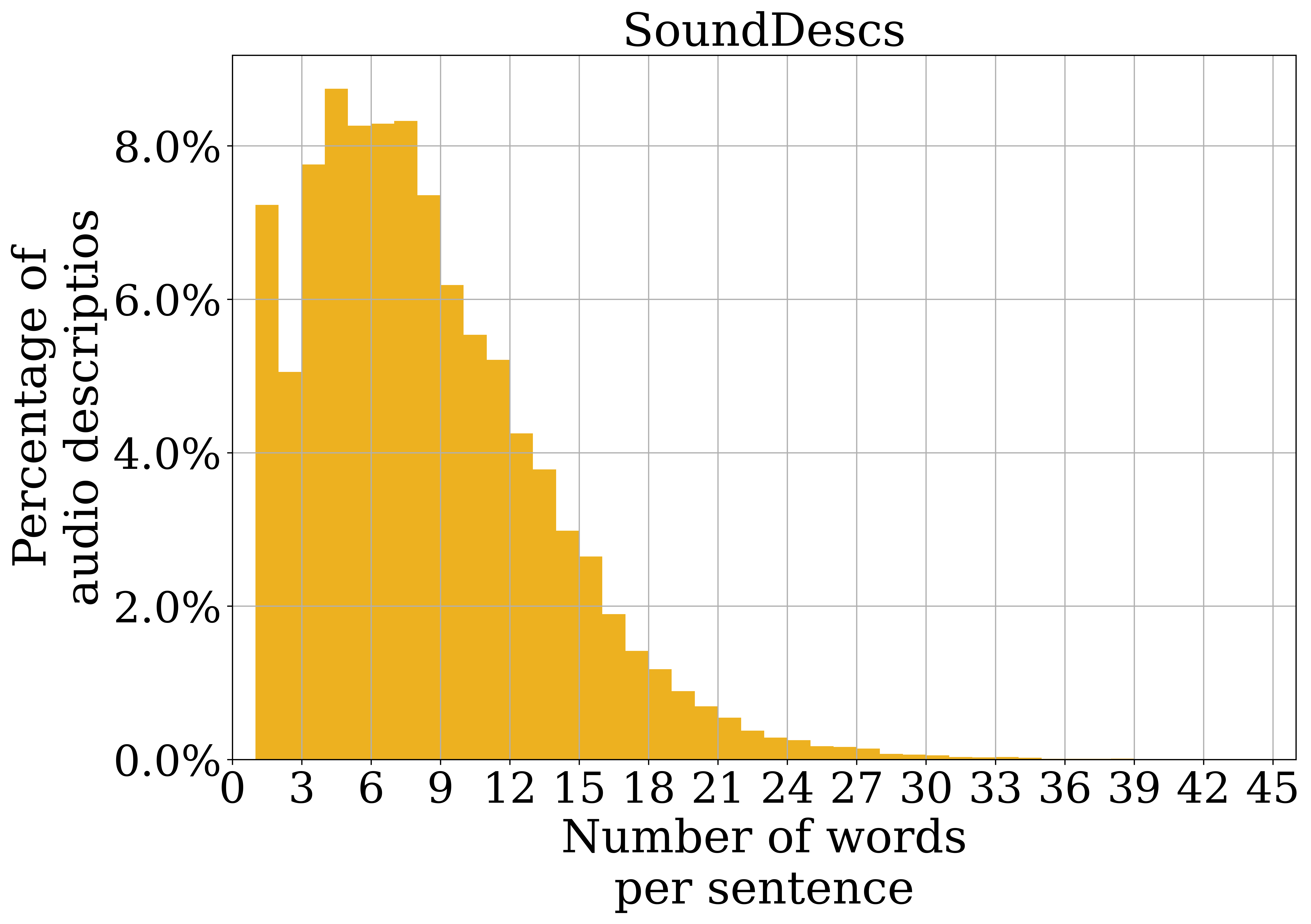}
    \caption{Sentence length distribution for \datasetName.}
    \label{fig:words_per_sent}
\end{figure}

To provide further insights into the text descriptions in the \datasetName dataset, Figures \ref{fig:nouns} and \ref{fig:verbs} show the distribution of unique nouns and verbs per description \rev{(0 unique nouns/verbs indicates that none were present)}. \datasetName contains noticeably more descriptions without any nouns and/or any verbs than \Audiocaps and \Clotho. There are many Nature instances in \datasetName which contain names of species of birds that are labelled as Proper Nouns instead of Nouns. Descriptions shown to not contain any verbs are either wrongly tagged, or they do not describe actions, e.g. \textit{Fountains: Rome - Sound of fountains, with street atmosphere}.
\begin{figure}[h]
    \centering
    \includegraphics[width=0.29\textwidth]{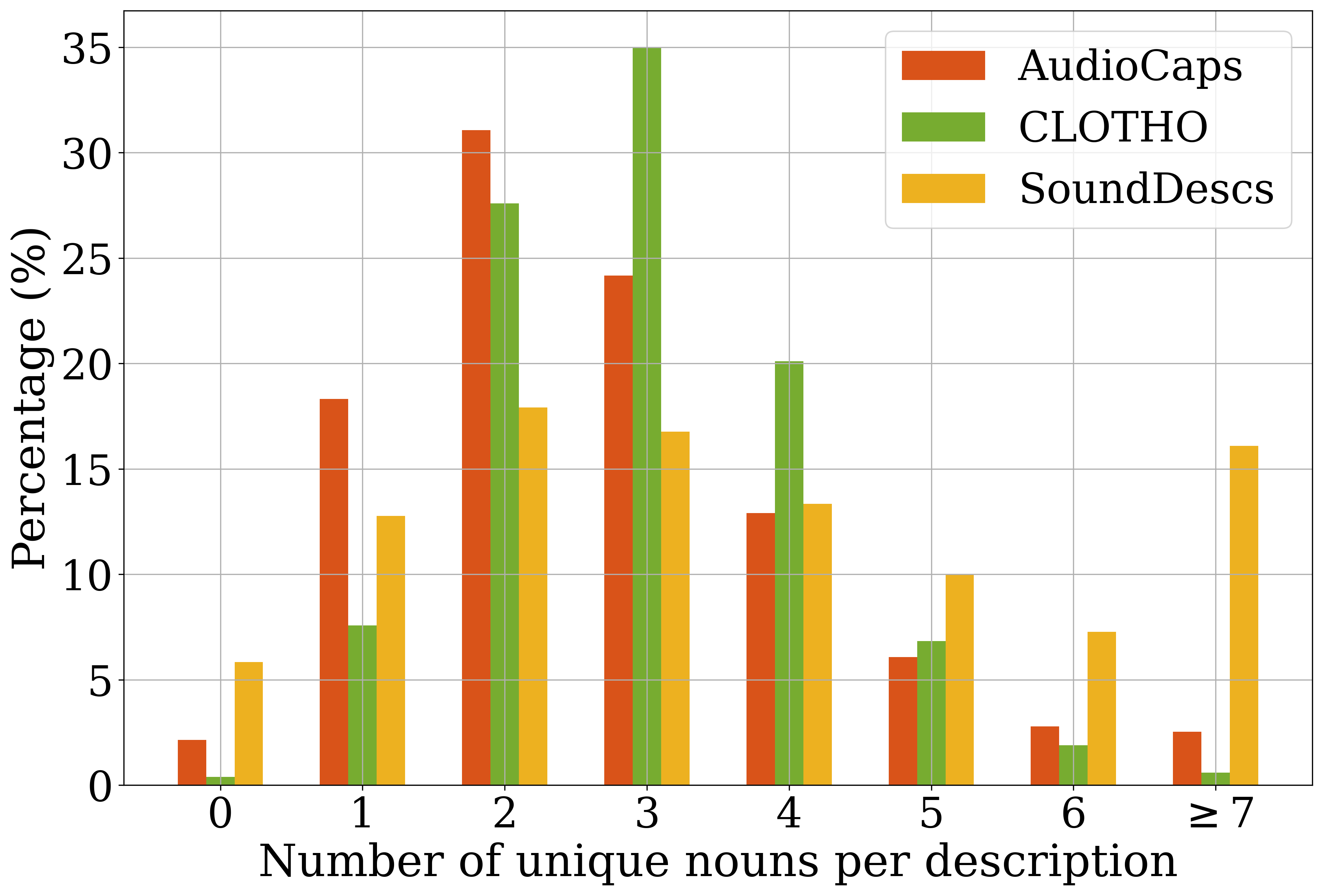}
    \caption{Distribution of unique nouns for descriptions in the \datasetName, \Audiocaps, and \Clotho datasets.}
    \label{fig:nouns}
\end{figure}
\begin{figure}[h]
    \centering
    \includegraphics[width=0.29\textwidth]{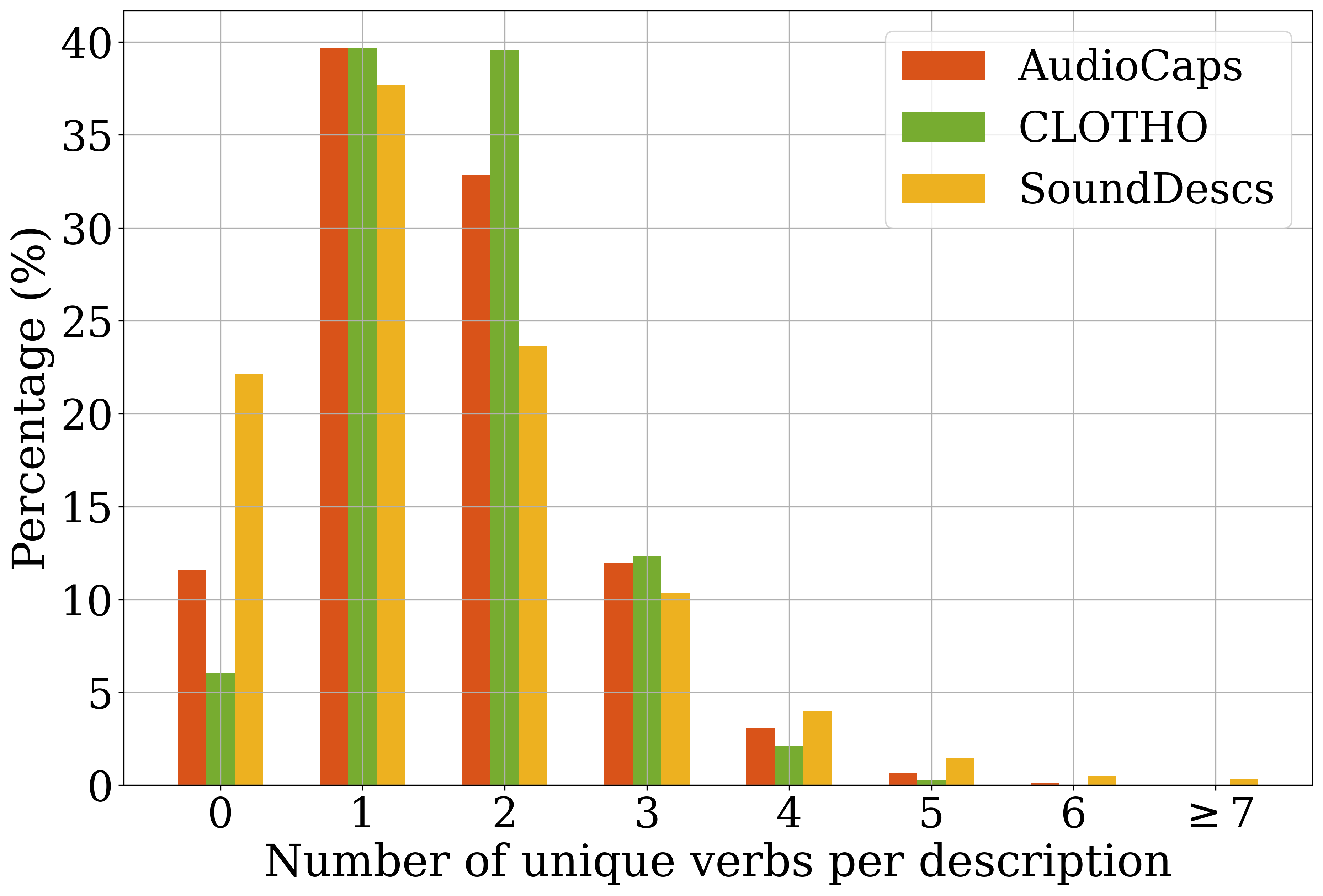}
    \caption{Distribution of unique verbs for descriptions in the \datasetName, \Audiocaps, and \Clotho datasets.}
    \label{fig:verbs}
\end{figure}

\end{section}

\ifCLASSOPTIONcaptionsoff
  \newpage
\fi

\bibliographystyle{IEEEtran}
\bibliography{vgg_bibtex/shortstrings,vgg_bibtex/vgg_local,vgg_bibtex/vgg_other,refs}

\end{document}